\documentclass[a4paper,11pt]{article}

\usepackage{jinstpub} 
\usepackage{subfigure}

\usepackage{verbatim}

\title{Radon Removal Commissioning of the PandaX-4T Cryogenic Distillation System}

\author[a,c]{Xiangyi Cui,}
\author[a,b,c,1]{Zhou Wang,\note{Corresponding author.}}
\author[f]{Jiafu Li,}
\author[b,g]{Shuaijie Li,}
\author[b]{Lin Si,}
\author[d,1]{Yonglin Ju}
\author[b]{Wenbo Ma,}
\author[a,b,c]{Jianglai Liu,}
\author[b,c]{Li Zhao,}
\author[e]{Xiangdong Ji,}
\author[d]{Rui Yan,}
\author[d]{Haidong Sha,}
\author[d]{Peiyao Huang,}
\author[d]{Xiuli Wang,}
\author[d]{Huaxuan Liu}

\affiliation[a]{Tsung-Dao Lee Institute, Shanghai Jiao Tong University, \\Shanghai 200240, China}
\affiliation[b]{School of Physics and Astronomy, Shanghai Jiao Tong University, Key Laboratory for Particle Astrophysics and Cosmology (MoE), Shanghai Key Laboratory for Particle Physics and Cosmology,\\Shanghai 200240, China}
\affiliation[c]{Shanghai Jiao Tong University Sichuan Research Institute,\\Chengdu 610000, China}
\affiliation[d]{Institute of Refrigeration and Cryogenics, Shanghai Jiao Tong University, \\Shanghai 200240, China}
\affiliation[e]{Department of Physics, University of Maryland, College Park,\\ Maryland 20742, USA}
\affiliation[f]{School of Physics, Sun Yat-Sen University, \\Guangzhou 510275, China}
\affiliation[g]{Yalong River Hydropower Development Company, \\Chengdu 610051, China}

\emailAdd{wangzhou0303@sjtu.edu.cn}
\emailAdd{yju@sjtu.edu.cn}

\abstract{
The PandaX-4T distillation system, designed for the removal of krypton and radon from xenon, is evaluated for its radon removal efficiency using a $^{222}$Rn source during the online distillation process. The PandaX-4T dark matter detector is employed to monitor the temporal evolution of radon activity.
To determine the radon reduction factor, the experimental data of radon atoms introduced into and bypassed the distillation system is compared. The results indicate that the PandaX-4T distillation system achieves a radon reduction factor exceeding 190 at the flow rate of 10~slpm and the reflux ratio of 1.44. Gas-only online distillation process of a flow rate of 20~slpm is also conducted without observing significant reduction of radon levels in the detector. This observation suggests that the migration flow of radon atoms from the liquid phase to the gas phase is limited, and the flow rate of gas circulation and duration of the process are insignificant compared
to the total xenon mass of 5.6~tons in the detector. 
This study provides the experimental data to support the efficient removal of radon at $\sim$Bq level using the PandaX-4T distillation system, which is the prerequisite of the radon background control in the detector. The further operation with higher flow rate will be applied for the upcoming science run in PandaX-4T.
}


\begin{document}
\maketitle
\flushbottom

\section{Introduction}
\label{sec:intro}

Liquid xenon has become a preferred choice for detecting ultra-low-rate signals due to its advantageous properties: it lacks long-lived radioisotopes except for $^{136}$Xe, has a large mass number, and can be maintained at a accessible cryogenic temperature. Recent advancements in multi-ton liquid xenon dual-phase time projection chamber (TPC) experiments, such as PandaX-4T~\cite{Panda4T}, LZ~\cite{LZ}, and XENONnT~\cite{xenonNT}, involve scaling up detector volumes to increase the probability of dark matter particles detection. However, achieving lower background levels in the detectors is crucial to enhance the sensitivity, which presents a significant challenge in terms of radioactivity control.

One of the key contributor to background radiation is $^{222}$Rn and its daughter isotope $^{214}$Bi with the $\beta$-emitter which influences electron recoil background in liquid xenon detectors. $^{222}$Rn is produced continuously within the detector materials from the decay of $^{238}$U and infiltrates the liquid xenon. Since radon atoms disperse evenly throughout the target volume, events induced by their decay cannot be eliminated through position selection during data analysis. The radon radioactivity levels in the experiments such as XENON1T, PandaX-4T, and LZ are all approximately 4~$\mu$Bq/kg~\cite{LZ,Panda4T,xenon1T}. However, even in the lowest activity experiment, XENONnT, radon still contributes one-third of background events roughly~\cite{xenonNT}.

Cryogenic distillation is one of the efficient methods to mitigate radon background, which is developed for krypton removal initially~\cite{XMass_kr1,panda2_kr1,panda2_kr2,xenon_kr}, has been adapted for radon suppression~\cite{boiling_rn,xenon100_rn,xenonnT_rn}, achieving a radon reduction factor exceeding 27 in XENON100 experiment. Additionally, absorption-based gas chromatography methods have been investigated for radon background control~\cite{xmass,LZ_Rn}.

In this paper, we focus on the PandaX-4T experiment containing approximately 5.6~tons of liquid xenon in China Jinping Underground Laboratory phase II (CJPL-II)~\cite{CJPL-II}. To reduce the intrinsic background radiation in the liquid xenon, the PandaX-4T distillation system was designed with two modes: krypton removal and a reversed operation process for radon removal~\cite{4T_dis}. The krypton removal process, which reached a purification rate of 10~kg/h and xenon collection efficiency of 99\%,  has purified 6~tons of xenon successfully, achieving a krypton concentration of 0.52~ppt (10$^{-12}$~mol/mol) in the liquid xenon of the TPC, which is consistent with the initial design. The distillation process separates radon, xenon, and krypton based on their different boiling points, enriching radon at the bottom of the distillation tower. The distillation system is designed to achieve a radon reduction factor of approximately $10^{3}$ while an expected reduction factor of 1.8 in the PandaX-4T detector at the circulation flow rate of 56.5~kg/h~\cite{4T_dis}.

In this paper, we present the results of the study conducted during the PandaX-4T commissioning run to assess the radon suppression effect by the distillation system under low flow rate conditions, limited by the circulation pump. The PandaX-4T distillation system is described briefly in section~\ref{sec:2}. The $^{222}$Rn injection method and events selection in the detector are discussed in section~\ref{sec:3}. The time evolution of radon activity during the radon distillation process and the results are discussed in section~\ref{sec:4}, before we conclude in section~\ref{sec:5}.

\section{PandaX-4T distillation system}
\label{sec:2}

The PandaX-4T distillation system comprises several main components, including the distillation packing column, the reboiler and the condenser, as well as the outer vacuum tower, automatic control system, the tube line system and the necessary supporting infrastructure. Based on the different boiling point between the xenon and radon (165~K and 211~K at atmosphere, respectively), the volatile component radon is enriched in the gas phase at the equilibrium state. 
The distillation process with multiple equilibrium states takes place inside the distillation packing column and the packing is used for enhancing the gas-liquid exchange and mass transfer area. The upstream gas is liquefied at the top condenser and parts of the gas are extracted as the Rn-free product. The ratio between the downstream liquid and the extracted product is called the reflux ratio, a key parameter affecting the enrichment ability. The downstream liquid xenon is stored in the bottom reboiler and parts of the liquid xenon is evaporated by the heater inside the reboiler. After the components exchange of the gas-liquid phase at the packings, a stable gradient of the radon concentration is built from the bottom reboiler to the top condenser during the stable operation.
The distillation packing column stands at a height of 6~m and contains 120 stainless steel structured packings with a diameter of 125~mm. The height equivalent of theoretical plate (indicates one equilibrium state) is considered as 350~mm by design, and the theoretical plates are calculated to be 17~\cite{4T_dis}. The stable operation temperature is 178~K and the condenser pressure is 2.2 atmospheric pressure (Bar) near the xenon saturation condition. The heating power of the reboiler is 120~W resulting in a reflux ratio of 1.44 and the gaseous xenon feeding rate is 10 standard liters per minute (slpm).


The improved McCabe-Thiele method is utilized in the design of the PandaX-4T distillation system~\cite{MT_method}
in order to achieve the reduction factors of krypton and radon in a single batch, which are 10$^{6}$ for krypton and 10$^3$ for radon, respectively. While the krypton impurity in commercial xenon gas used for PandaX-4T could be effectively removed using cryogenic distillation before xenon is injected into the detector, the radon removal process requires a different approach to involve circulation with the detector, referred as online radon distillation.

To assess the radon removal efficiency, two important parameters are considered: the distillation reduction factor denoted as $R_{\rm Dis}$, and the detector reduction factor labeled as $R_{\rm Det}$. $R_{\rm Dis}$ reflects the radon removal ability of the distillation system, and the factors such as distillation tower construction, radon emanation rates from the packing materials, and the initial radon concentration of the inlet xenon are considered. However, $R_{\rm Det}$ relies primarily on the circulation flow rate $f$, especially when $R_{\rm Dis}$ significantly exceeds 1, as illustrated in the equation~\ref{eq:2:1}~\cite{4T_dis}.
\begin{equation}
    \centering
    \label{eq:2:1}
    R_{\rm Det} = \frac{\lambda _{\rm Rn} + f \cdot \left( 1 - 1/R_{\rm Dis} \right)}{\lambda _{\rm Rn}} ,
\end{equation}
where $\lambda_{\rm Rn}=2.1\times10^{-6}$~s$^{-1}$ is the radon decay constant.

\begin{figure}[htbp]
\centering 
\includegraphics[width=.96\textwidth]{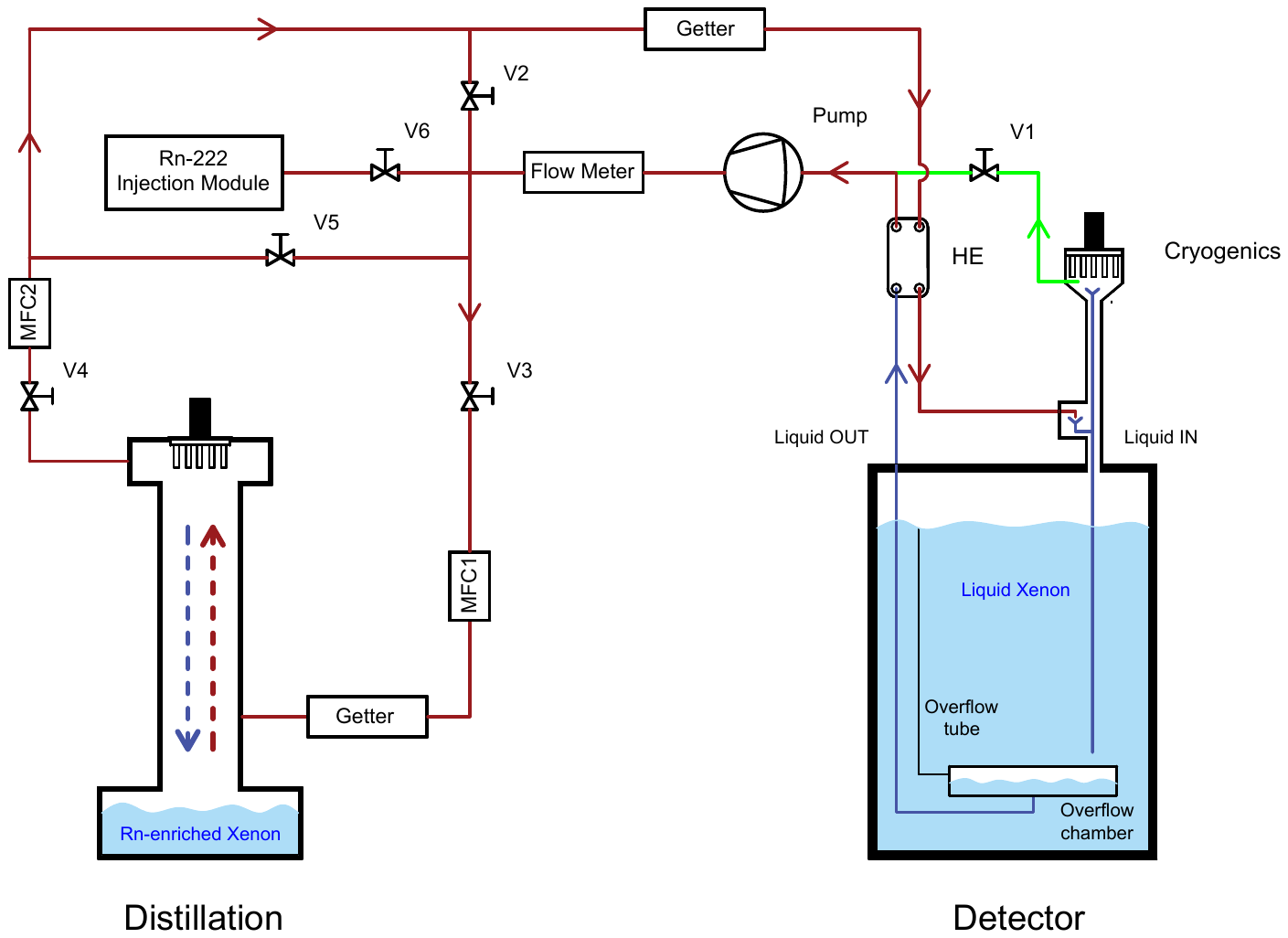}
\caption {The P\&I diagram of the PandaX-4T distillation system and its interfaces with the detector. Red lines are the shared loop for the liquid and gas online distillation, blue and green lines indicate the liquid-only and gas-only online distillation, respectively.}
\label{fig:2:1}
\end{figure}

A simplified piping and instrument (P\&I) diagram of the distillation system integrated with the PandaX-4T detector is illustrated in Figure~\ref{fig:2:1}. During the online radon distillation process, liquid xenon is extracted from the overflow chamber at bottom of the TPC to maintain a balanced liquid level. This extracted liquid xenon undergoes a phase transformation to gaseous in the heat exchanger (HE) with the xenon gas from the circulation system. Subsequently, it passes through the circulation pump and the mass flow meter~\cite{zhaoli}. After being purified by the high temperature getter, the xenon is injected into the packing tower at 13th theoretical plate number counted from the top. The distillation processes unfold within the packing tower, leading to the enrichment of radon in the reboiler located at the bottom of the tower, in which radon is left to undergo self-decay in this study.

During the stable operation, a consistent radon concentration gradient is maintained from the bottom to the top. The gaseous product xenon in the condenser at the top of the distillation tower flows back into the circulation system due to pressure differential, and purified by the high-temperature getter. Subsequently, majority of the gaseous xenon is liquefied by the HE, then flows back into the detector. The cycle continues as the liquid xenon is drawn from the TPC via the HE by the circulation pump, and returns to the distillation tower. The distillation flow rate is controlled by a mass flow controller (MFC), limited by the ability of the circulation pump (pressure limitation of 3.5~Bar) and the flow resistance of the system. The pressures of the distillation tower and the circulation system are monitored by pressure gauges. In addition to the liquid xenon circulation for online distillation, gaseous xenon could also be extracted from the cryogenics system called gas-only distillation. Furthermore, two distillation bypasses are integrated, including the activation of valve $V2$ in Figure~\ref{fig:2:1} within the circulation loop and valve $V5$ in Figure~\ref{fig:2:1} within the distillation control panel, which contains the tube lines of approximately 40~m.

Assuming there is no Rn-enriched xenon extracted from the reboiler, the operation equations in the rectifying and stripping section could be modified as
\begin{subequations}
\label{eq:4:2}
\begin{align}
\label{eq:4:2:1}
{\rm Rectifying\ Section:\ } &y_{\rm Rn} = \frac{R}{R+1} \cdot x_{\rm Rn} + \frac{c_{D}}{R+1} ,
\\
\label{eq:4:2:2}
{\rm Stripping\ Section:\ } &y_{\rm Rn} =  x_{\rm Rn} - \frac{c_{B} \cdot \lambda_{\rm Rn} \cdot m_{\rm Xe,reb.}}{R \cdot D},
\end{align}
\end{subequations}
where $x_{\rm Rn}$ and $y_{\rm Rn}$ are the radon concentration in the liquid and gas phase, respectively. $R$ is the reflux ratio and $D$ is the product xenon flow rate extracted from the condenser along with the radon concentration $c_{D}$. The radon concentration in the reboiler $c_{B}$ is related to the xenon mass in the reboiler $m_{\rm Xe,reb.}$ that described as $c_{B} \cdot \lambda_{\rm Rn} \cdot m_{\rm Xe,reb.} = F \cdot c_{F} - D \cdot c_{D}$~\cite{MT_method,xenonnT_rn} where $F$ and $c_{F}$ are the flow rate and the radon concentration in the injected xenon from the detector.

Based on the operation equations, the expected radon reduction factor of the PandaX-4T distillation tower is more than 10$^{3}$ at the operation conditions applied in this study.

The radon removal ability of the distillation system is also affected by the emanated radon from the packings as mentioned in Ref.~\cite{4T_dis}. During the PandaX-4T commissioning run, the decay of $^{222}$Rn with and without the online krypton distillation is 8.6~$\mu$Bq/kg and 6.0~$\mu$Bq/kg, respectively~\cite{Panda4T}, which means the maximum radon emanation rate of each theoretical plate is about 1.5$\times$10$^{-27}$~mol/s/stage, a value that could be ignored in this study.

\section{Radon injection and events selection}
\label{sec:3}
In order to obtain the removal ability of the distillation system for ultra-low radon concentration less than 10$^{-20}$~mol/mol, the PandaX-4T detector is utilized to monitor the temporal evolution of the radon concentration. To provide more precise evaluation of the distillation reduction factor, an external $^{222}$Rn source is introduced upstream of the distillation tower, under the assumption that the expected radon reduction factor of the distillation system exceeds 10$^3$.

\subsection{$^{222}$Rn source}
\label{sec:3:1}

\begin{figure}[htbp]
\centering 
\subfigure[$^{226}$Ra particles, with the 1/4-inch VCR gasket]{
\label{fig:3:1:a}
\begin{minipage}[t]{0.45\textwidth}
\centering
\includegraphics[width=5.6cm]{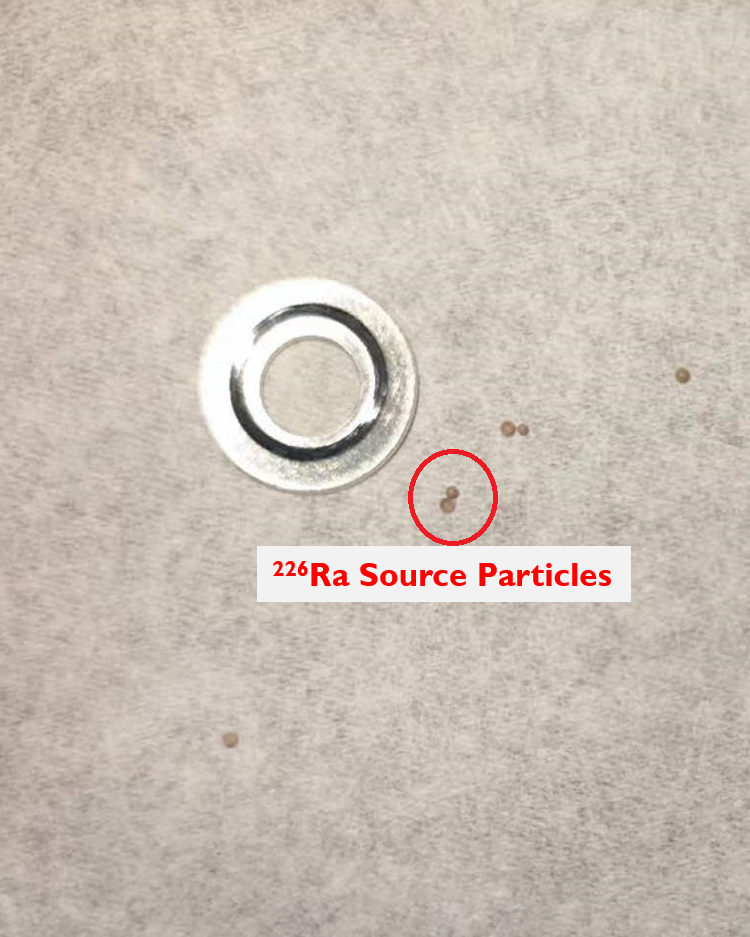}
\end{minipage}
}
\subfigure[Injection module in front of the control panel]{
\label{fig:3:1:b}
\begin{minipage}[t]{0.45\textwidth}
\centering
\includegraphics[width=5.6cm]{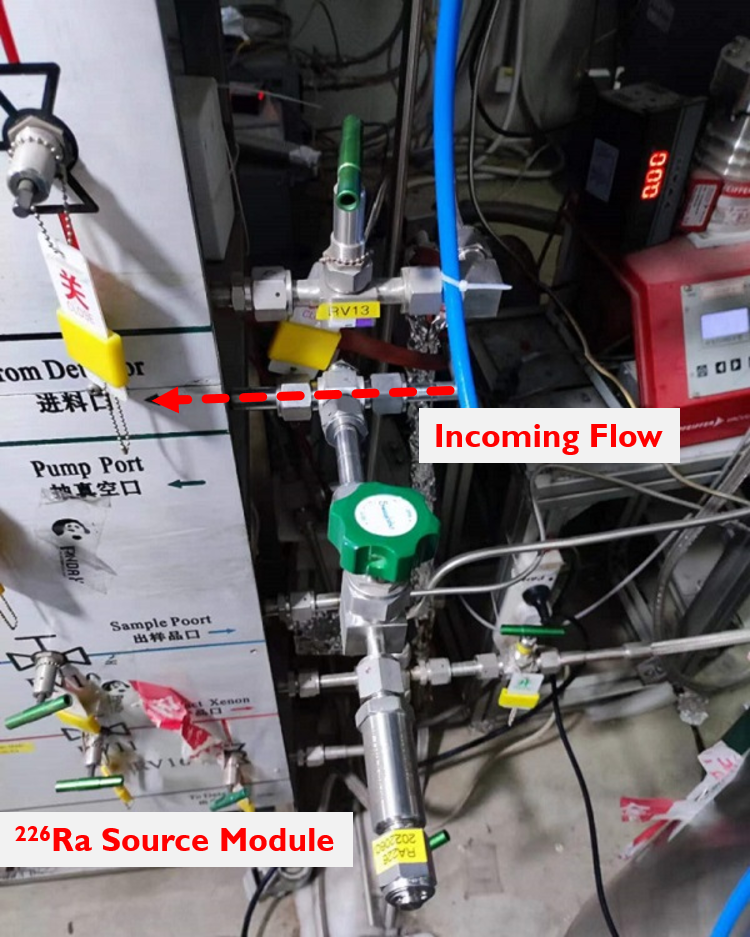}
\end{minipage}
}
\caption{Photos of the $^{226}$Ra source particle and the injection module.}
 \label{fig:3:1}
\end{figure}

$^{222}$Rn is the decay daughter of $^{226}$Ra with a half-life of 1.6 years. In this study, the spheroplast granular $^{226}$Ra source from the University of South China is used to generate $^{222}$Rn atoms~\cite{source1,source2}. Representations of the $^{226}$Ra source particles and the injection module are presented in Figure~\ref{fig:3:1}. To prevent potential dust contamination of the source to the detector, a 3-nanometer particle filter is positioned between the source and the sealed valve. 
The diffusion method is employed. Different from injecting radon atoms to the xenon gas directly, the radon source installed at the branch of the mainstream as shown in Figure~\ref{fig:3:1:b}, so that the radon atoms are diffused to the injection tube then mixed with the incoming flow. 
The radon radioactivity of each $^{226}$Ra source particle is 4~Bq, measured by a commercial RAD7 device~\cite{RAD7}, and 10 particles are placed.

\subsection{Radon events selection}
\label{sec:3:2}

The PandaX-4T experiment utilizes a dual-phase liquid xenon time projection chamber technique, as outlined in Ref.~\cite{Panda4T}. The deposited energy of events in the liquid xenon releases the prompt scintillation photons ($S1$) and ionized electrons. The ionized electrons undergo drift towards to the top of the TPC under the electric field and then extracted from the liquid phase to the gas phase through a stronger extraction field. Through the process of electron luminescence, the ionized electrons will released a delayed scintillation photons ($S2$). The $S1$ and $S2$ signals are collected by the top and bottom photomultiplier tubes (PMT). 

\begin{figure}[htbp]
        \centering 
        \subfigure[S1TBA versus qS1 before the alpha events selection, where dashed black lines are the selection criteria.]{
        \label{fig:3:2:a}
        \begin{minipage}[t]{0.48\textwidth}
        \centering
        \includegraphics[width=8cm]{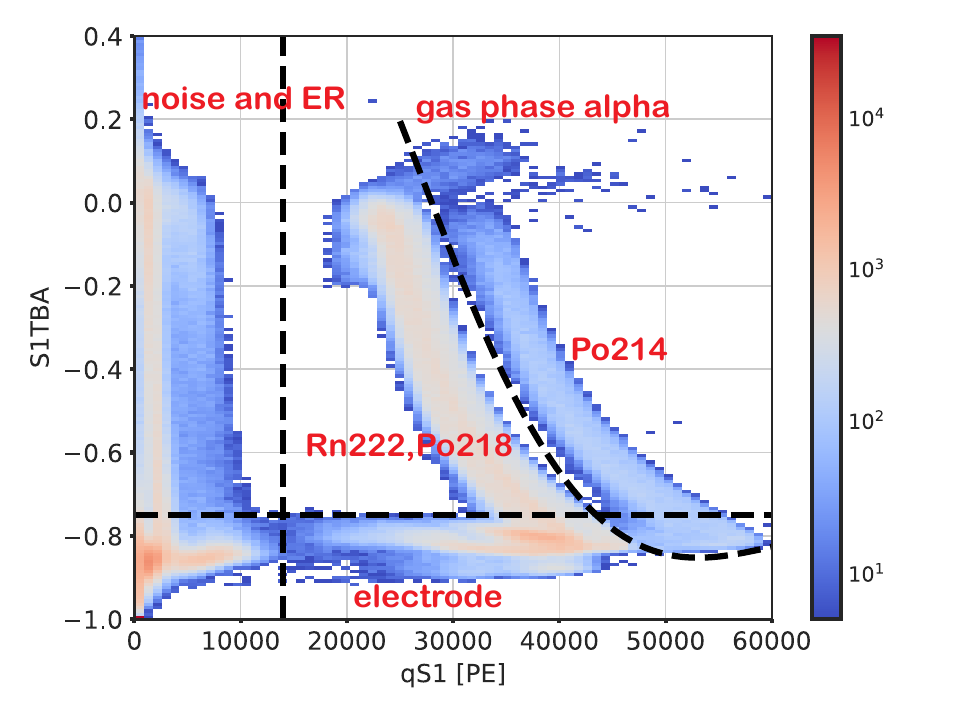}
        \end{minipage}
        }
        \subfigure[S1TBA versus hS1 after the alpha events selection.]{
        \label{fig:3:2:b}
        \begin{minipage}[t]{0.48\textwidth}
        \centering
        \includegraphics[width=8cm]{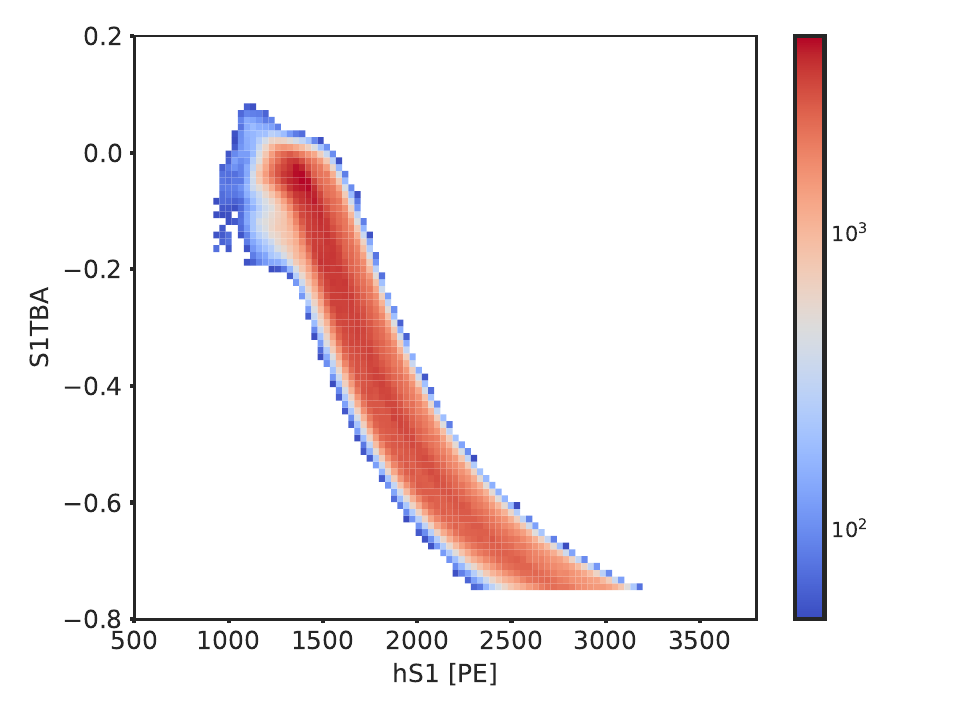}
        \end{minipage}
        }
        \caption{ \label{fig:3:2} The events distribution with S1-only information in PandaX-4T.}
        \end{figure}

The $S1$ and $S2$ waveform will be stored and processed a few sequential steps to reconstruct the incoming event. The horizontal position ($x,y$) of the event is reconstructed from the collected photons distribution of the $S2$ signal in the top PMT arrays. And the vertical position ($z$) of the event can be derived from the time difference between the $S1$ and $S2$ signals. In addition, the ratio between the $S2$ and $S1$ for the nuclear recoil (NR) signal produced by the dark matter scattering with xenon nucleus is significantly lower than the electron recoil (ER) background produced by the gamma or beta events. 
The alpha events by the radon decays can be easily distinguished from other backgrounds due to the high energy and large $S1$-to-$S2$ ratio with the much higher electron recombination ratio~\cite{Shaoli}. Figure~\ref{fig:3:2:a} shows the $S$1TBA calculated by (qS1T-qS1B)/qS1 versus $S1$ amplitude (qS1) events distribution in the PandaX-4T detector, where the qS1T and qS1B are the $S1$ amplitude observed by the top and bottom PMTs, respectively.

\begin{figure}[htbp]
            \centering
            \includegraphics[width=.9\textwidth]{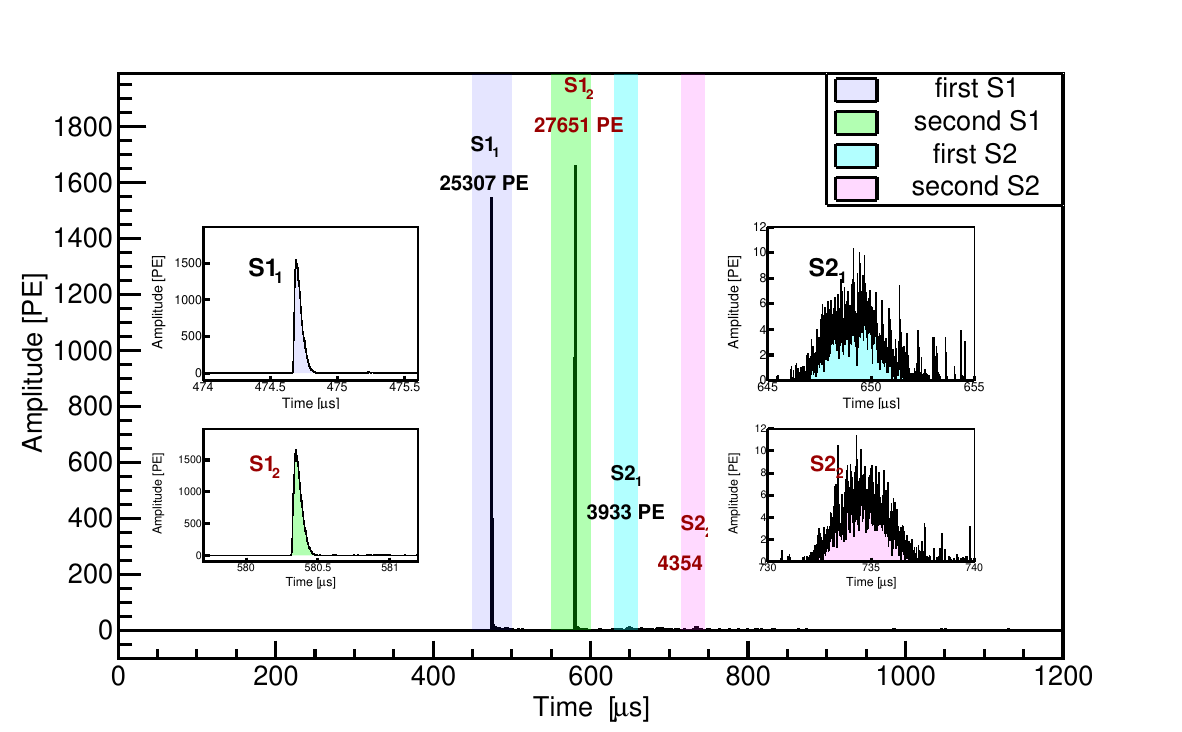}
            \caption{Waveform of two radon signals occurring within the same time window in the PandaX-4T detector.}
            \label{fig:3:3}
        \end{figure}

        A typical single scattering event like the conventional dark matter signal is characterized by the presence of one $S1$ pulse accompanied by one associated $S2$ pulse. So, one paired $S1$ and $S2$ pulse are classified and rebuilt the incoming event in the event selection time window which spans 1 millisecond. In this study, we did not require the single scattering condition as the multiple events would be recorded in the same time window with the high radon activity. Figure~\ref{fig:3:3} displays two instances of radon $S1$ signals occurring in the same time window, where the qS1 requirement exceeds 1.3$\times$10$^4$ photonelectrons (PE).

        $^{222}$Rn emits a 5.5~MeV alpha particle during its decay to $^{218}$Po, followed by another alpha decay to $^{214}$Pb with the energy of 6.0~MeV with the half-life of 3.1 minutes. Approximately 3.5×10$^4$~PE of $S1$ signal is expected to be observed for the alpha event originating from $^{222}$Rn in the PandaX-4T detector, based on the photon detection efficiency (PDE) of about 9.0\%~\cite{Panda4T}. The PDE remains stable during the injection period which is related to the detector structure and the impurity of the liquid xenon. However, the $S2$ signal is associated with the electron extraction efficiency (EEE) influenced by the gate voltage, which means the $S2$ signal has a risk to be mis-identified once the gate voltage is not high enough. According to that, only the $S1$ information, including qS1, S1 waveform height (hS1) and the $S$1TBA, was required in this analysis, called the $S1$-only analysis. As the collected photons by the top and bottom PMT arrays are changed according to the vertical position of the physical event, the $S$1TBA is relevant to the event scattering position in the liquid xenon, which could reject events from the noise and the electrode discharges. 
        In summary, the alpha selection criteria in this study is hS1>400~PE and the area is shown in Figure~\ref{fig:3:2:a}, which is between qS1>1.3$\times 10^4$~PE, $S$1TBA>-0.75 and the left edge of the $^{214}$Po band.
        According to the alpha selection criteria, the sensitive volume can be treated as the candidate selection region which equals to 3.7~tonne~\cite{Panda4T}. More detailed description of the signal process in PandaX-4T can be seen in Ref.~\cite{4T-signal}. Figure~\ref{fig:3:2:b} presents the distribution of $S$1TBA versus hS1 after the alpha event selection criteria.

        Without the $S2$ information,the scattering position of the radon event is hardly distinguished that the $S1$ spatial uniformity correction in the detector is challenging. In this case, the $^{222}$Rn and $^{218}$Po events could not be separated easily. Half of the events passing the alpha events selection criteria are attributed to the $^{222}$Rn contribution to address this limitation, and the specific ratio for this attribution is determined through an examination of $S1$-$S2$ pairing events.

\section{Online radon distillation}
\label{sec:4}

In this study, approximately 30~kg of commercial xenon is introduced into the distillation tower to initiate the total reflux process, achieving a state of self-equilibrium, which establishes a stable distribution of xenon and radon compositions. Subsequently, two distinct processes are needed to executed in order to assess the radon reduction efficiency of the PandaX-4T distillation system: 1. The TPC circulation involves the injection of radon atoms passing through the distillation tower; 2. The TPC circulation involves the injection of radon atoms bypassing the distillation system. 

During the experiment, krypton, oxygen, and other lighter components are enriched in the condenser, while the heavier radon component is accumulated in the reboiler. Consequently, the detector would be contaminated potentially by lighter impurities when the radon-free xenon extracted from the condenser, and it is essential to eliminate these impurities especially for the initial total reflux establishment with commercial xenon. Several recuperation processes, xenon extracted from the condenser and restored in an independent bottle, are employed until the krypton concentration reaches approximately 0.4~ppb (4$\times$10$^{-10}$~mol/mol)~\cite{kr_meas}, in which level the contribution of krypton contamination becomes negligible for the PandaX-4T detector.

The experiment involves four stages based on different operation conditions:

    \textbf{(1) Initial Radon Injection:} Radon is injected initially bypass the distillation tower at the circulation flow rate of 10~slpm. After 8 hours injection, a radon activity of 25~Bq is achieved in the system.

    \textbf{(2) Self-Circulation Without Radon Injection:} Following the initial injection, the detector undergoes self-circulation without any radon injection for approximately 12~days.

    \textbf{(3) Second Radon Injection:} The second radon injection consists of two steps—first passing through the distillation tower at a flow rate of 10~slpm and then bypassing it. Each step lasts approximately 4~days, with the maximum radon activity in the detector reaching about 18~Bq.

    \textbf{(4) Gas-only Online Distillation:} In this step, gas-only online distillation is performed at a flow rate of 20~slpm without any radon injection for around 5~days.
    

\subsection{Initial radon injection}
\label{sec:4:1}

The radon injection module along with the interface tubes is evacuated to a vacuum level with an out-leakage rate of less than 10$^{-12}$ Pa·m$^3$/s measured by helium mass spectrometer. Prior to the radon injection, the radon concentration increases from 48.7$\pm$0.6~mBq to 81.1$\pm$1.0~mBq after connecting the circulation loop. This emanation is anticipated to originate from various instruments including the getter and the KNF pump.

After connecting the interface tubes of the injection module to the detector, the radon activity increases to 474.2$\pm$2.2~mBq, which is ten times higher compared to the TPC-only background. This increase is suspected due to the contamination in the interface tubes of radon atoms, which have a length of approximately 40~m. This contamination occurs during the vacuum pumping period probably and diffuses into the xenon gas during circulation subsequently.

\begin{figure}[htbp]
\centering 
\includegraphics[width=.96\textwidth]{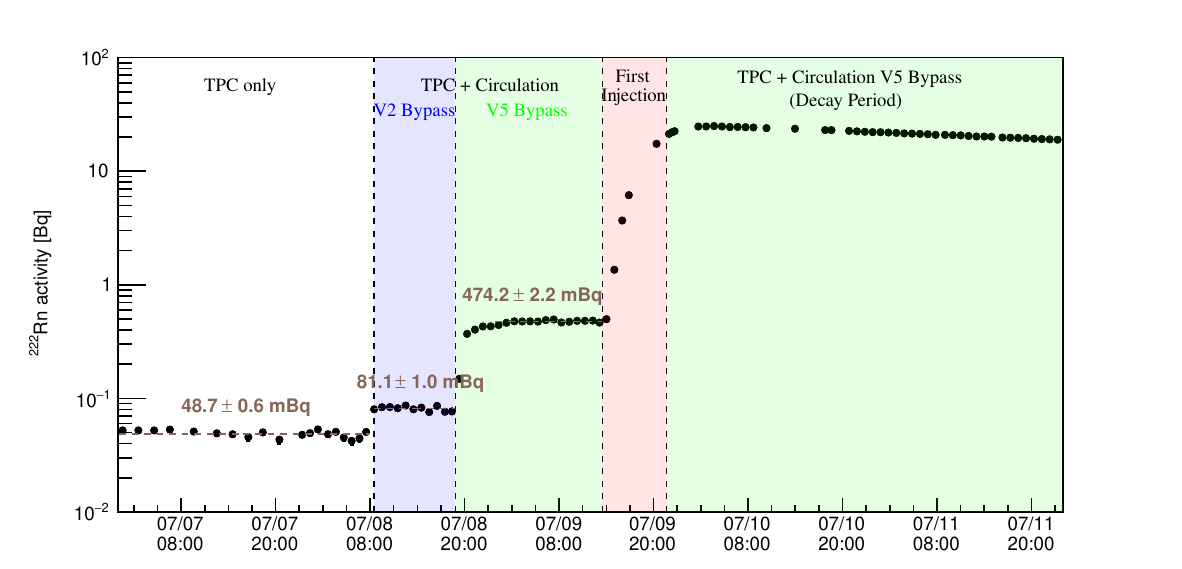}
\caption{\label{fig:4:2} {Time evolution of the $^{222}$Rn activity during the first injection campaign, with fitted radon activity values before the radon injection.}}
\end{figure}

The first radon injection campaign commences and continues for approximately 8~hours until the gate voltage trips. Following the isolation of the radon source, the $^{222}$Rn rate continues to increase for a few hours. The maximum radon activity measured in the detector reaches 25~Bq. Figure~\ref{fig:4:2} illustrates the time evolution of the radon activity during the first injection campaign.

After the first injection campaign, the detector is self-circulated and waiting for the radon self-decay. To avoid such high Rn activity, 4/5 of $^{226}$Ra source particles are removed, then the injection module is pumped independently during this time.

\subsection{Self-circulation without radon injection}
\label{sec:4:2}

\begin{figure}[htbp]
\centering 
\includegraphics[width=.96\textwidth]{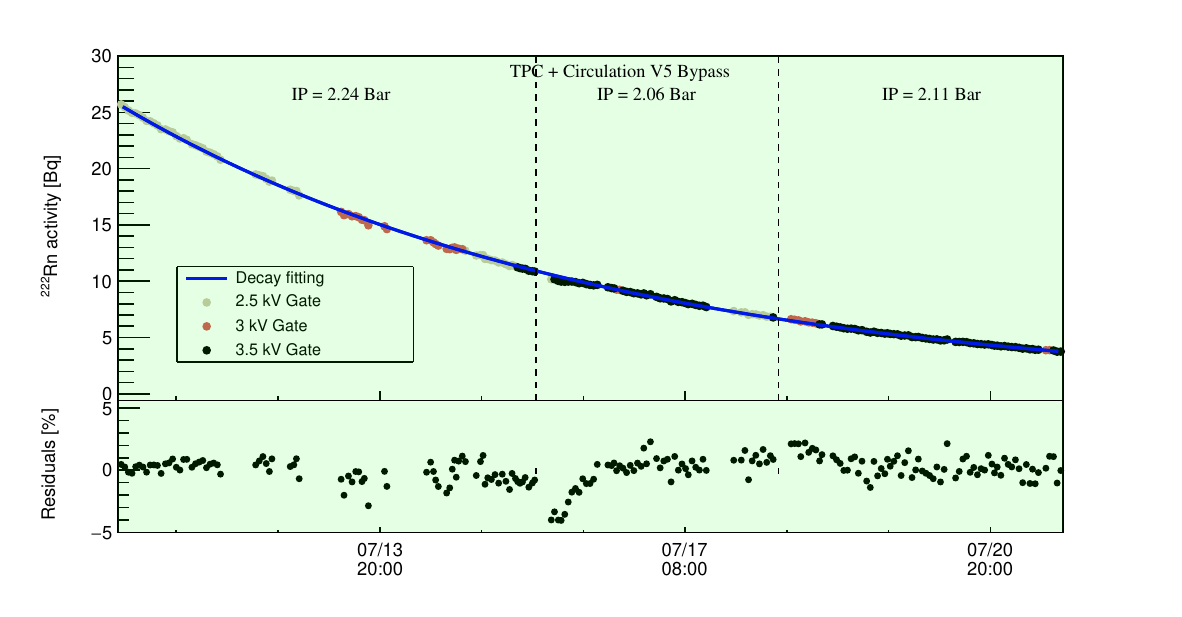}
\caption{\label{fig:4:3} {Time evolution of the $^{222}$Rn activity after the first injection with the residuals calculated by (Data-Fit)/Data. The dots with different color represent different gate voltages and the solid blue line is the fitting curve.}}
\end{figure}

A radon activity of $\sim$4~Bq is achieved in the detector after 12~days of self-circulation. The gate voltage is operated under three different conditions during the decay period: 2.5~kV, 3~kV, and 3.5~kV, respectively. To ensure the stability of the gate voltage, the inner pressure (IP) of the xenon gas in the detector is adjusted twice transitioning from 2.24~Bar to 2.11~Bar.
Figure~\ref{fig:4:3} illustrates the time evolution of the $^{222}$Rn activity following the initial injection. As depicted in the figure, the selection efficiency of radon events remained consistent across different gate voltage settings.

The time evolution of the radon atoms $dN(t)/dt$ in the detector is described as the differential equation:
\begin{equation}
\centering
\label{eq:4:2:1}
\frac{dN(t)}{dt} = k_{0} - \lambda_{\rm fit} \cdot N(t) ,
\end{equation}
where $k_{0}$ denotes the baseline of the radon activity emanated from the detector itself, and $\lambda_{\rm fit}$ is the fitted radon decay constant affected by different operation modes.

Because the actual baseline during the online distillation process is hardly to be obtained, there are three platform values before the first injection period are tested to describe the decay trend, including $k_{0}$ values of 48.7~mBq, 81.1~mBq and 474.2~mBq.
However, only one fitting curve is presented in Figure~\ref{fig:4:3} for simplification.
Considering the baselines of the TPC alone (48.7~mBq) and the addition of the circulation loop (81.1~mBq), the best-fitting decay constant is determined to be (2.08$\pm 0.01)\times$10$^{-6}$~s$^{-1}$, which closely matches the expected decay constant of 2.1$\times$10$^{-6}$~s$^{-1}$.
Considering all the interface tubes between the distillation unit and the detector as a baseline of 474.2~mBq, the best-fitting decay constant of the decay tendency is (2.16$\pm 0.01)$$ \times$10$^{-6}$~s$^{-1}$.

\subsection{Second radon injection}
\label{sec:4:3}

The second injection campaign involves in two distinct processes: one in which the circulating xenon gas contains the radon source passing through the distillation system, and another in which the xenon bypasses the distillation system entirely. Liquid-only circulation with a flow rate of 10~slpm is employed lasting approximately 4~days for each process.

\begin{figure}[htbp]
\centering 
\includegraphics[width=.96\textwidth]{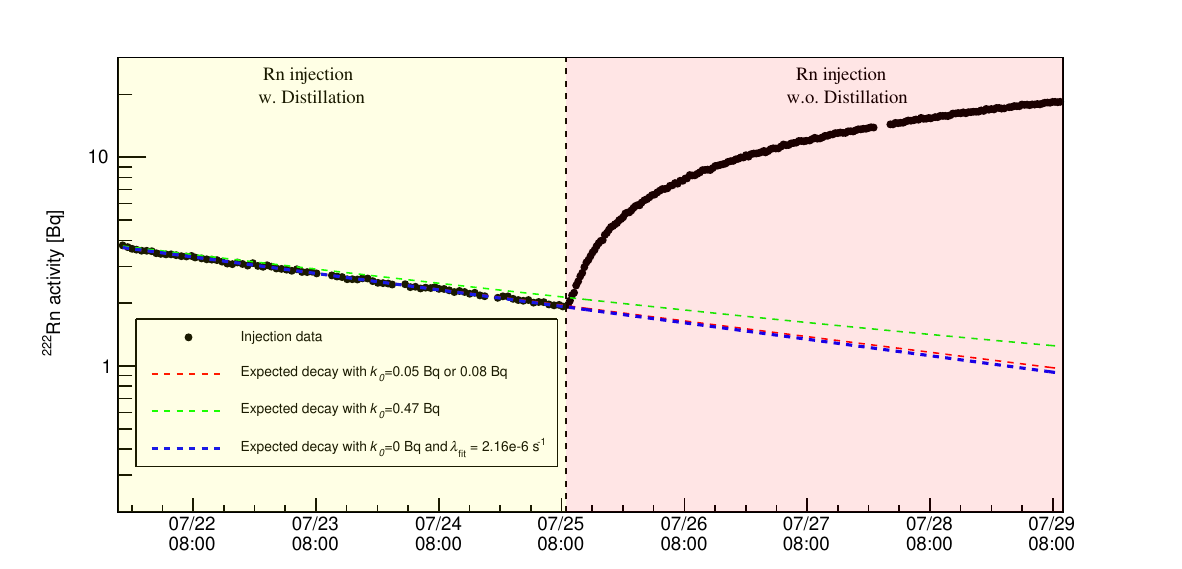}
\caption{\label{fig:4:4} {Time evolution of the $^{222}$Rn activity during the second injection campaign, where the red and green dashed lines are the expected decay tendency with different baseline assumptions from the data of the decay period. The blue dashed line is the decay curve with the baseline of 0~Bq and the decay constant of 2.16$\times$10$^{-6}$~s$^{-1}$.}}
\end{figure}

When radon atoms passes through the distillation tower during the injection, the radon activity in the detector decreases to about 2~Bq gradually. 
Subsequently, the radon activity begins to increase when the radon bypasses the distillation tower, which means the distillation system is isolated with the xenon flowing through the V5 and the V3\&V4 are closed in Figure~\ref{fig:2:1}.
Figure~\ref{fig:4:4} displays the time evolution of radon activity in the detector during the second injection campaign.
In this figure, it demonstrates various baseline assumptions during the decay period, resulting in different decay tendencies, which impact the reduction calculation subsequently.
The radon time evolution is calculated via the experimental data minus the expected decay tendency, which means the quicker decay tendency brings the higher radon activity. The decay tendency would be quick with the assumption of lower baseline or shorter decay constant. In this case, more radon atoms survive after the distillation so that the result of $R_{\rm Dis}$ is more conservative.
To adopt the most conservative approach for estimating the expected decay tendency, a baseline of 0~Bq is assumed and a decay constant of 2.16$\times$10$^{-6}$~s$^{-1}$ is got from the fitting of the decay period with the 474.2~mBq baseline assumption.

\begin{figure}[htbp]
\centering 
\includegraphics[width=.96\textwidth]{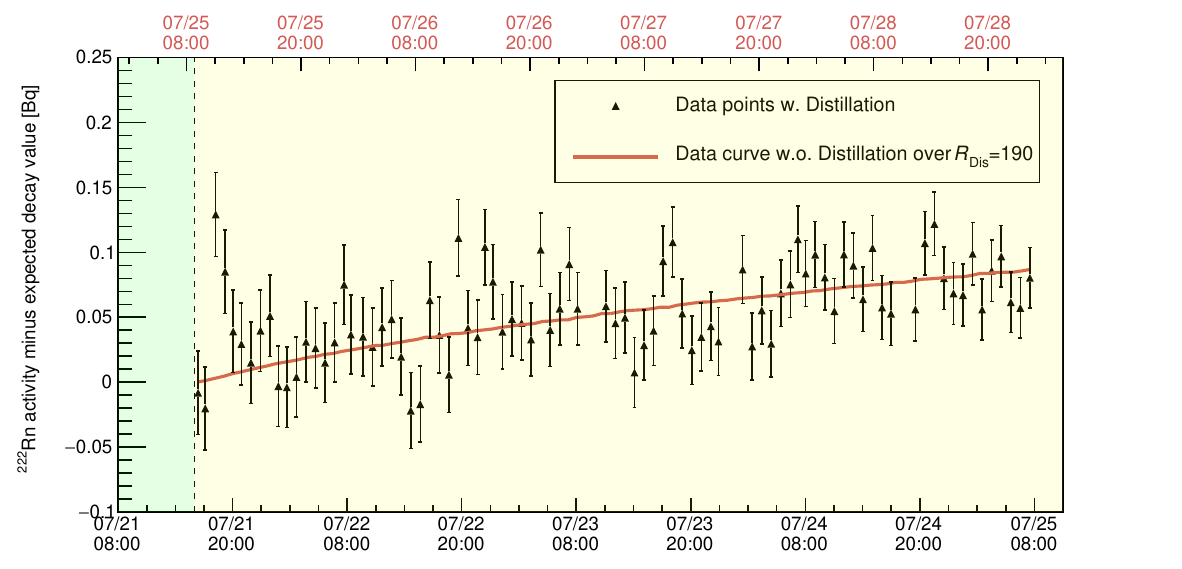}
\caption{\label{fig:4:5} {$^{222}$Rn activity during the radon injection passing the distillation minus the expected decaying value, where the dashed blank line indicates the injection start time and the solid red line is the fitted curve.}}
\end{figure}

To calculate $R_{\rm Dis}$, the data from the period of radon bypassing the distillation tower is utilized to fit the radon injection activity when it passes through the distillation tower. We assume that the increasing tendencies in the two injection campaigns are similar due to the identical system construction. Figure~\ref{fig:4:5} illustrates the radon time evolution of the radon atoms passing through the distillation tower, minus the expected decaying tendency, along with statistical error bars. The red line represents the fitted curve with the expression $g_{\rm w.o.}/R_{\rm Dis}$, where $g_{\rm w.o.}$ represents the radon injection data bypassing the distillation tower minus the expected decaying tendency. The fitting yielded a radon reduction factor of {$R_{\rm Dis} = (1.9\pm0.1)$$\times10^2$} of the PandaX-4T distillation tower, which is within the expected reduction range. According to equation~\ref{eq:2:1}, the $R_{\rm Det}$ equals to 1.08 at this experimental condition.

\subsection{Gas-only online distillation}
\label{sec:4:4}

After the second injection, the gas-only online radon distillation process is operated for 5~days at a flow rate of 20~slpm. The total xenon mass in the gas phase of the TPC is approximately 20~kg, and the radon atoms in the gas phase are efficiently removed during the online radon distillation process. As previously discussed in other studies~\cite{xenon_KrAr,xenon_O2}, the radon atoms tend to migrate from the liquid phase to the gas phase continuously to establish equilibrium, which is advantageous for radon control within the detector.

\begin{figure}[htbp]
\centering 
\includegraphics[width=.96\textwidth]{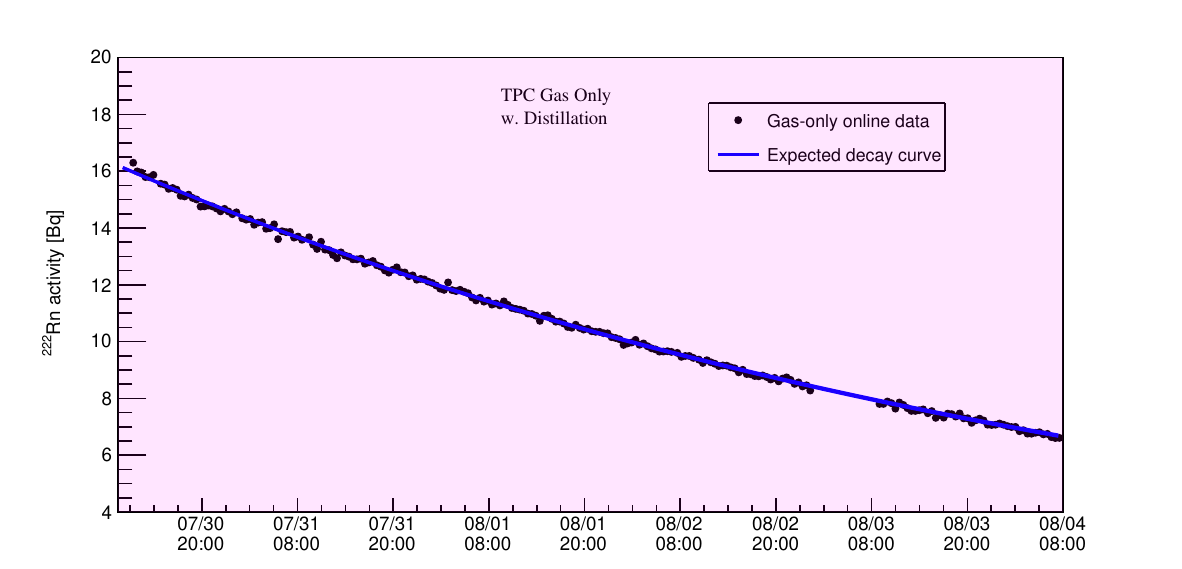}
\caption{\label{fig:4:6} {The time evolution of the $^{222}$Rn activity during the gas-only online distillation, where the blue solid line is the expected decay curve from the data of decay period. }}
\end{figure}

Figure~\ref{fig:4:6} displays the time evolution of radon activity during the gas-only online distillation and there is no significant evidence of the radon suppression in the detector was observed. Several factors could contribute to this result, including a relatively low circulation flow rate, a short distillation time that may not have allowed for the establishment of radon distribution equilibrium in the TPC, as well as the potentially limited efficiency in the diffusion of radon from the liquid phase to the gas phase. 
Furthermore, the radon suppression method of the gaseous xenon component at room temperature need to be considered because the radon releasing rate from the materials is faster at room temperature than at cryogenic temperature~\cite{lz_arxiv}.
Future deeper radon removal studies of the cryogenic distillation system are anticipated to involve longer observation periods exceeding two weeks, and higher circulation flow rates reaching 160~slpm.

\section{Conclusion}
\label{sec:5}

The radon removal ability of PandaX-4T distillation is investigated using a $^{222}$Rn source. The study involves the integration of the PandaX-4T detector, the circulation system, and the cryogenic distillation system. Online radon distillation is successfully conducted at a maximum flow rate of 20~slpm. The PandaX-4T detector is utilized to monitor the time evolution of radon activity, and the scintillation signal analysis is developed for this purpose.
Comparing the two radon injection processes of passing through the distillation system and bypassing it, a radon reduction factor of 190 is obtained for the cryogenic distillation system. 
The radon suppression in the detector is mainly related to the online distillation flow rate under this $R_{\rm Dis}$ value. With the operation of the online distillation mode and other ongoing improvements, the flow rate of 160~slpm is expected to be achieved, leading the radon level to reduce below 3~$\mu$Bq/kg in the next PandaX-4T dark matter run.

\acknowledgments

The authors would like to thank the supports of the PandaX-4T collaboration. We also thank the radon source application from the University of South China.
This project is supported by grants from the Ministry of Science and Technology of China (No. 2016YFA0400301 and 2016YFA0400302), a Double Top-class grant from Shanghai Jiao Tong University, grants from National Science Foundation of China (Nos. 11435008, 11505112, 11525522, 11775142, 11755001, 12205189 and 52206015), grants from the Office of Science and Technology, Shanghai Municipal Government (Nos. 11DZ2260700, 16DZ2260200, and 18JC1410200), and the support from the Key Laboratory for Particle Physics, Astrophysics and Cosmology, Ministry of Education. 
We also thank the sponsorship from the Chinese Academy of Sciences Center for Excellence in Particle Physics (CCEPP), Hongwen Foundation in Hong Kong, and Tencent Foundation in China. Finally, we thank the CJPL administration and the Yalong River Hydropower Development Company Ltd. for indispensable logistical support and other help.


\end{document}